\documentclass{article}
\usepackage{spconf,amsmath,epsfig}
\usepackage{graphicx}
\usepackage{amsfonts}

\graphicspath{ {./images/} }

\title{Minimizing Estimation Error Variance using a Weighted Sum of Samples from the Soil Moisture Active Passive (SMAP) Satellite}
\name{Mohammad Koosha, Nicholas Mastronarde\thanks{The work of M. Koosha and N. Mastronarde was supported in part by the NSF under Award \#2030157.}}
\address{University at Buffalo, The State University of New York}

\begin{document}
%
\maketitle
\begin{abstract}
The National Aeronautics and Space Administration's (NASA) Soil Moisture Active Passive (SMAP) is the latest passive remote sensing satellite operating in the protected L-band spectrum from 1.400 to 1.427 GHz. SMAP provides global-scale soil moisture images with point-wise passive scanning of the earth's thermal radiations. SMAP takes multiple samples in frequency and time from each antenna footprint to increase the likelihood of capturing RFI-free samples. SMAP's current RFI detection and mitigation algorithm excludes samples detected to be RFI-contaminated and averages the remaining samples. But this approach can be less effective for harsh RFI environments, where  RFI contamination is present in all or a large number of samples. In this paper, we investigate a bias-free weighted sum of samples estimator, where the weights can be computed based on the RFI's statistical properties. 
\end{abstract}
\section{Introduction}
\label{sec:intro}
Passive radiometry is the crucial foundation for passive remote sensing with a wide range of applications from climatology to earth sciences. Empirical field experiments and data from multiple passive remote sensing satellites indicate
increasing Radio Frequency Interference (RFI) across the globe, despite international radio regulations for preserving protected RFI-free passive sensing frequency bands \cite{alam2022radio}. This alarming increase in the RFI level experienced by passive radiometry satellites calls for improvements in RFI mitigation techniques. 
\vspace{3pt}

Although satellite-borne passive remote sensing has been done in various frequency bands, including the C-band and X-band, the new generation of passive radiometry satellites, starting with the European Space Agency's (ESA) Soil Moisture and Ocean Salinity (SMOS) satellite \cite{kerr2001soil}, operate in the protected portion of the L-band from 1.400 to 1.427 GHz. The National Aeronautics and Space Administration's (NASA) Soil Moisture Active Passive (SMAP) is one of the latest in a series of remote sensing satellites that provides high-resolution images of global soil moisture with passive scanning of earth's thermal radiations in this band \cite{entekhabi2014smap}.  
\vspace{3pt}

SMAP's digital backend takes multiple measurements for each antenna footprint and sends them to Earth for ground-based processing. An RFI detection and mitigation algorithm 
removes samples detected as being RFI-contaminated from the pool of samples of each antenna footprint, and averages the remaining samples \cite{bringer2021properties}. Since no sample can be precisely declared RFI-free, we investigate a probabilistic approach where each sample has an assigned RFI distribution parameterized by its mean and variance.
In contrast to SMAP's approach,
we investigate a weighted sum of samples where each sample is assigned a weight. Assuming the means and covariance matrix of the RFI values are known,
we model the estimation error variance as a quadratic function and calculate the weight for each sample. We show that this estimator is bias-free and minimizes the estimation error variance.
\vspace{3pt}

The paper is organized as follows: Section \ref{sec:prems} outlines SMAP's radiometer design and RFI detection algorithms. Section \ref{sec:wsvr} presents our proposed RFI minimization method. Section \ref{sec:sims} showcases simulation results. Section \ref{sec:con} concludes and discusses future work.


\section{Preliminaries}
\label{sec:prems}

\begin{figure*}[!ht]
  \includegraphics[width=\linewidth]{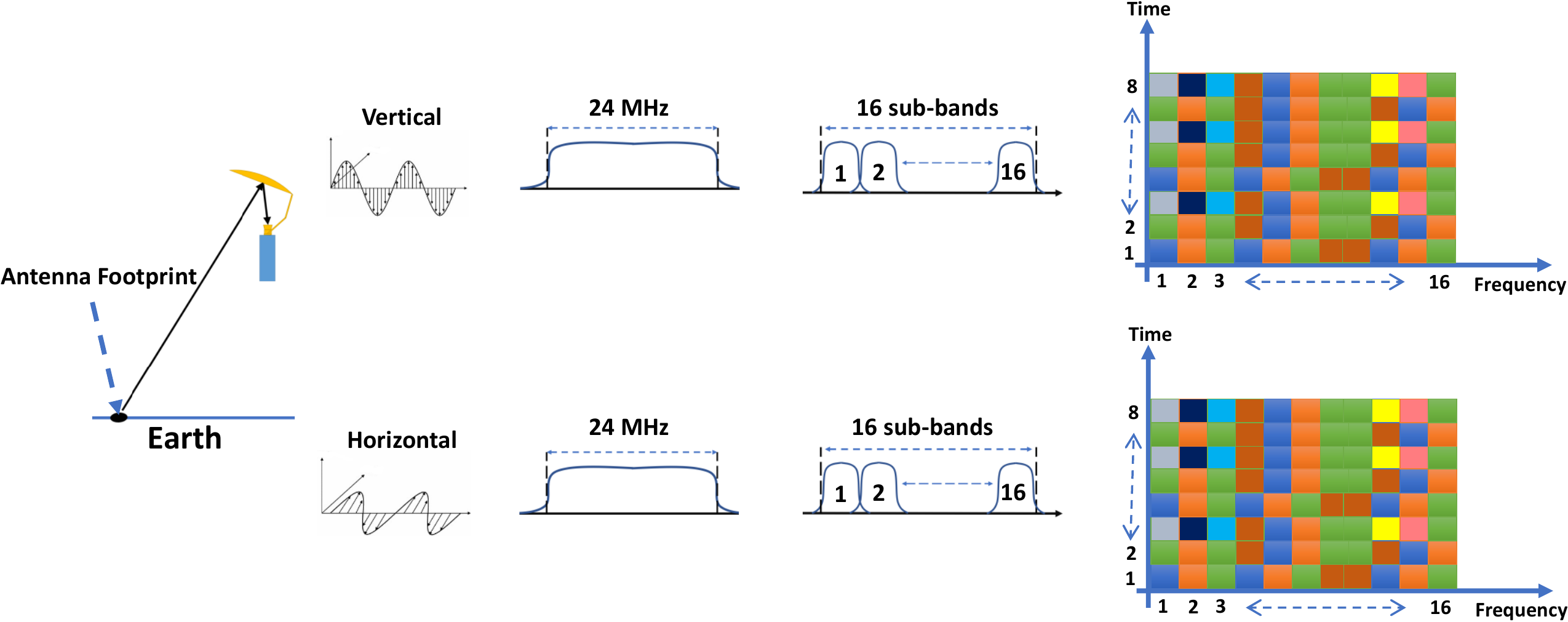}
  \caption{SMAP's digital back-end}
  \label{pic1}
\end{figure*}

SMAP captures 
high-resolution images of earth's soil-moisture content with periodic global coverage every two to three days \cite{koosha2022opportunistic}. SMAP has a 6-meter-wide conically-scanning golden mesh reflector with a 3-dB antenna beam width of $2.4^{\circ}$ that projects a footprint of roughly $40 \times 40$ km$^2$ on earth with an incident angle of approximately $40^{\circ}$. 
An Ortho-Mode Transducer (OMT) feedhorn collects the reflected radiations from the mesh reflector and duplexes them separately into vertical and horizontal polarizations \cite{piepmeier2016soil} (Fig. \ref{pic1}).
\vspace{3pt}

In the next phase, for each polarization, SMAP's Radiometer Digital Electronics (RDE) sample the received signal in a 24 MHz full-band channel centered at 1.413 GHz. The sampled signal then
goes through a poly-phase filter bank with 16 sub-bands of 1.5 MHz. The RDE provides 8-time samples of 1.2 ms integration time for each antenna footprint at each subband. As illustrated in Fig. \ref{pic1}, the final products are two $16 \times 8$ frequency-time spectrograms for both vertical and horizontal polarizations \cite{piepmeier2016soil}. 
This sampling strategy increases the probability that SMAP captures RFI-free samples under sporadic interference in both frequency and time.
\vspace{3pt}

SMAP's superiority is its advanced Digital Signal Processing (DSP) backend, which enables ground-based science RFI detection and removal algorithms. Along with frequency-time spectrograms, SMAP provides  $2^{nd}$, $3^{rd}$, and $4^{th}$ raw moments of both full-band and frequency-time samples of spectrograms, along with the $3^{rd}$ and $4^{th}$ stokes parameters of both sub-bands and full-bands. This auxiliary information provides the basis for rendering anomaly detection algorithms on the received data and excluding potentially RFI-contaminated frequency-time samples \cite{piepmeier2016soil}. 
\vspace{3pt}

Each frequency-time sample has to go through 9 different RFI detection algorithms of pulse, cross-frequency, kurtosis, and polarization-anomaly types. According to \cite{bringer2021properties}, most of these 9 RFI detection methods operate by comparing the deviation of a particular measurement $T$ from its expected mean $m$ to a threshold, where the threshold is a multiple $\beta$ of the expected standard deviation $\sigma$ of the measurement, with detection declared for measurements satisfying 
\begin{equation}
    |T-m| \geq \beta \sigma \label{eqq9}.
\end{equation}
The aggregate detection techniques can detect narrow-band, wide-band pulse, and continuous wave RFI. Each RFI detection algorithm assigns each frequency-time sample an RFI flag. In a maximum Probability of Detection (PD) approach, each frequency-time sample has to pass all 9 RFI detection algorithms without any ``on'' RFI flag. In the last phase, the frequency-time samples declared RFI-free are averaged \cite{bringer2021properties}.
\vspace{3pt}

In recent years, advanced techniques for detecting RFI in SMAP sampling structures have emerged. For instance, Alam et al. \cite{alam2022radio} proposed a cutting-edge supervised deep learning method utilizing convolutional neural networks. This method effectively identifies RFI-contaminated pixels in SMAP's frequency-time spectrograms, providing a promising alternative to traditional detection approaches. However, existing literature predominantly focuses on RFI detection rather than optimizing the recovery and estimation of soil moisture.

\section{Proposed Method}
\label{sec:wsvr}

Most RFI detection techniques used by ground-based science algorithms rigidly throw away samples above specific predefined threshold values, and the remaining samples are crudely averaged. This approach is suboptimal from an estimation theory perspective because
(a) the final product is a biased estimate and (b) the error variance of the final estimate is not minimized. These two problems arise because most samples cannot be precisely declared RFI-free, especially those that are marginally below or above the threshold values used by ground-based RFI detection techniques. As noted in \cite{mohammed2021microwave}, RFI appears as a random additive value in each sample, possibly with a Generalized Extreme Value (GEV) probability density function (pdf). Instead of rigidly excluding samples and averaging the remaining samples, we investigate a weighted sum of samples approach where the sample weights are determined according to the RFI's mean and covariance matrix.

As noted in Section \ref{sec:prems}, SMAP provides a total of $2 \times 16 \times 8 =256$ samples from each antenna footprint.  Each sample $p_i$ can be expressed as the sum of the earth's thermal emissions in the antenna's footprint ($p_{soil}$ Watts) and an additive RFI random variable ($P^{i}_{RFI}$ Watts) with mean $\mu_{i}$, i.e.,
\begin{align}
p_{i}=P_{RFI}^{i}+p_{soil} \quad \text{Watts}, \quad i=1,..,256 \label{eq1}.
\end{align}
%
We write the measured samples in vector form as $\textbf{p}=[p_{1},...,p_{i},...,p_{256}]^{T}$, the RFI experienced by each sample in vector form as $\textbf{P}_{RFI}=[P^{1}_{RFI},...,P^{i}_{RFI},...,P^{256}_{RFI}]^{T}$, and the mean RFI for each sample in vector form as $\boldsymbol \mu=[\mu_{1},...,\mu_{i},...,\mu_{256}]^{T}$. We assign each sample $p_{i}$ a weight of $\alpha_{i}$ to make a weighted sum average. We write the weight vector as $\boldsymbol A =[\alpha_{1},...,\alpha_{i}, ..., \alpha_{256} ]^{T}$. 
We define $\Bar{p}$ as an unbiased estimator of $p_{soil}$ using a weighted sum of samples:
%
%
\begin{align}
    \Bar{p}=\boldsymbol A^{T} (\boldsymbol p - \boldsymbol \mu) &= \sum_{i=1}^{256} \alpha_{i} (p_{i}-\mu_{i}) \label{eqq2} \\
    &=\sum_{i=1}^{256} \alpha_{i} (P_{RFI}^{i}-\mu_{i})+\sum_{i=1}^{256}\alpha_i p_{soil}  \notag \\
    &=\boldsymbol A^{T} (\boldsymbol P_{RFI} - \boldsymbol \mu)+ p_{soil} \label{eqq3} \\
    \text{subject to:}\quad \boldsymbol 1^{T} \boldsymbol A=1, \notag 
\end{align}
where $\boldsymbol 1 = [1,...,1]^{T}$ is a 256-element vector of ones. The estimator $\Bar{p}$ in \eqref{eqq2} is an unbiased estimator since $\mathbb{E}[\Bar{p}]=p_{soil}$. To increase the accuracy of estimator $\Bar{p}$ in \eqref{eq1}, we must minimize the \textbf{estimation error} $Err=\Bar{p}-p_{soil}$. Since $\mathbb{E}[Err]=0$, smaller values of the estimation error variance $\sigma^2_{Err}$ means that error values are concentrated closer to zero. Using \eqref{eqq3} we write the \textbf{estimation error variance} $\sigma^2_{Err}$ as follows:
\begin{align}
    \sigma_{Err}^{2}&=\mathbb{E} \big [ (\Bar{p}-p_{soil})^{2} \label{eqq5} \big] \\
    &= \mathbb{E} \big [ \boldsymbol A^{T}(\boldsymbol P_{RFI} - \boldsymbol \mu) (\boldsymbol P_{RFI} - \boldsymbol \mu)^{T} \boldsymbol A \big] \notag \\
    &= \boldsymbol A^{T} \mathbb{E}\big [(\boldsymbol P_{RFI} - \boldsymbol \mu) (\boldsymbol P_{RFI} - \boldsymbol \mu)^{T}\big] \boldsymbol A \notag \\
    &= \boldsymbol A^{T} \boldsymbol \Sigma \boldsymbol A \label{eqq6}
\end{align}
where $\boldsymbol \Sigma$ is the covariance matrix of $\boldsymbol P_{RFI}$, i.e., 
\begin{align}
\boldsymbol \Sigma=\begin{bmatrix}
    \sigma_{1,1}^2 & \sigma_{1,2}^2 & \hdots & \sigma_{1,256}^2 \\ 
    \sigma_{2,1}^2 & \sigma_{2,2}^2 & \hdots & \sigma_{2,256}^2 \\ 
    \vdots & \vdots  & \vdots & \vdots \\
    \sigma_{256,1}^2 & \hdots &\hdots & \sigma_{256,256}^2 
\end{bmatrix} \notag.
\end{align}
To minimize the estimation error variance, the sample weight vector $\boldsymbol A$ must be selected to minimize $\sigma^2_{Err}$. In \eqref{eqq6}, $\sigma^2_{Err}$ is a quadratic function, and the weights vector $\boldsymbol A$ can be acquired using Quadratic Programming (QP) convex optimization \cite{fletcher1971general}. The general problem can be stated as follows:
\begin{align}
   &\underset{\boldsymbol A}{\arg\min} \quad \boldsymbol A^{T} \boldsymbol \Sigma \boldsymbol A \label{eqq7} \\
   &\text{subject to:}\quad \boldsymbol 1^{T} \boldsymbol A=1 \notag.
\end{align}
Equation \eqref{eqq7} has a solution for positive definite (PD) matrix $\boldsymbol \Sigma$. For the particular case of diagonal $\boldsymbol \Sigma$, the problem turns into a Mean Square Error (MSE) problem \cite{rousset2019reducing}. However, a diagonal covariance matrix means that RFI contaminating samples are completely independent; thus, off-diagonal  elements are all zero. Using Lagrange multiplier $\lambda$, we can write the general solution to \eqref{eqq7} as \eqref{eqq8} \cite{fletcher1971general}: 
\begin{align}
    \begin{bmatrix}
    \boldsymbol \Sigma & \textbf{1} \\
    \textbf{1}^{T} & 0
\end{bmatrix}
\begin{bmatrix}
    \boldsymbol A \\
    \lambda
\end{bmatrix} =
\begin{bmatrix}
    0 \\
    \vdots \\
    0 \\
    1
\end{bmatrix}. \label{eqq8}
\end{align}



\section{Simulation Results}
\label{sec:sims}

\begin{figure}
  \centering
  \includegraphics[width=0.9\linewidth]{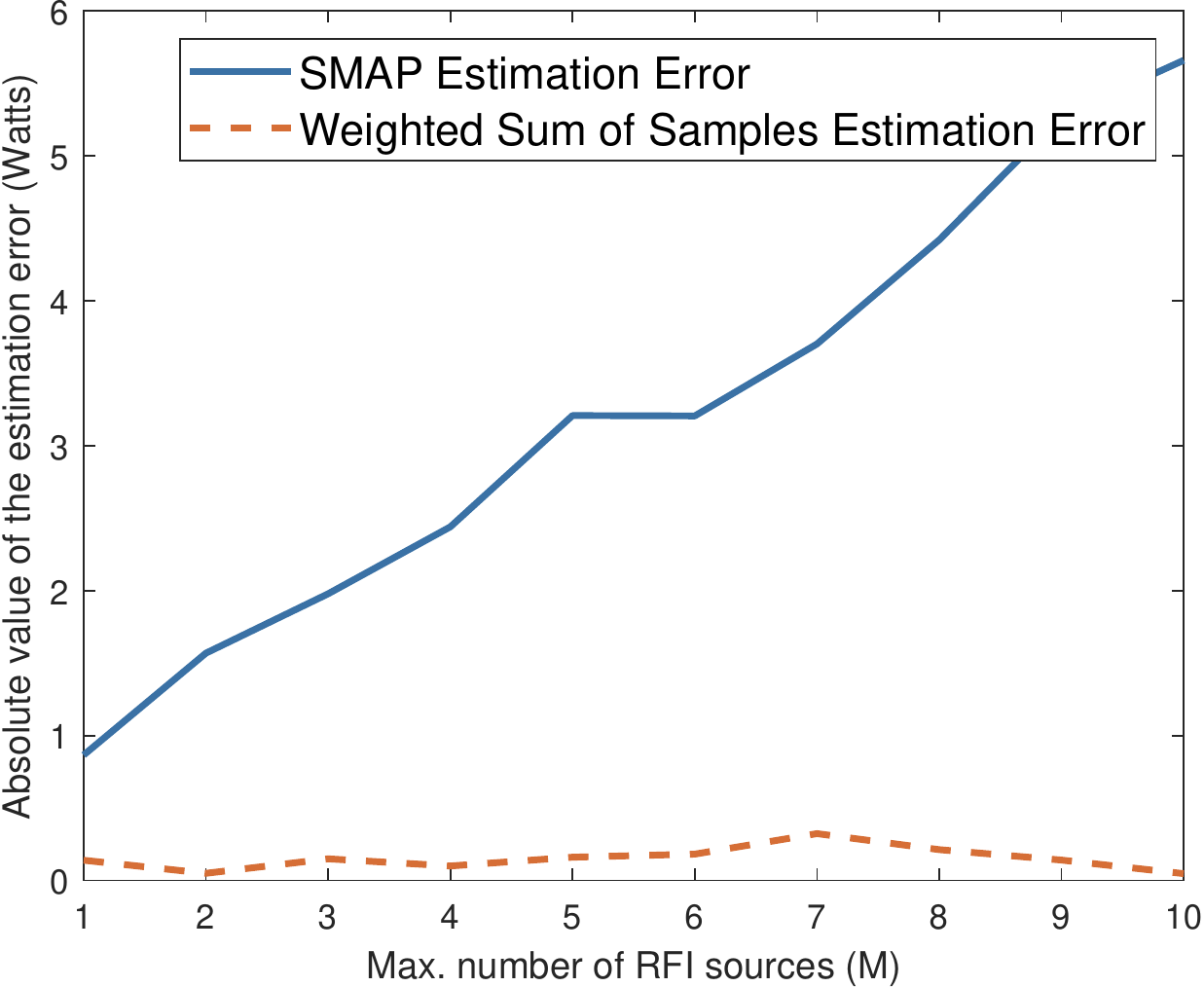}
  \caption{Comparison of estimation errors.}
  \label{Err}
\end{figure}

In this section, we examine the performance of our proposed weighted sum of samples method compared to SMAP's RFI mitigation technique in terms of estimation error and estimation error variance.
\vspace{2pt}

For illustration, we imagine a maximum of $M$ RFI sources in the environment. Each RFI source has independent sporadic emissions both in time and frequency. Consequently, each SMAP sample $p_i$ can be impacted by $k_i = 1,2,\ldots, M$ RFI sources.
For each sample $p_i$, $k_i$ is randomly and 
uniformly chosen between 1 and $M$. Each RFI source impacting sample $p_i$ has a normally distributed amplitude. Consequently,  $P_{RFI}^{i}$ (power in Watts) is the sum of $k_i$ independent squared normal random variables, i.e., it has a chi-squared distribution $\chi^{2} (k_i)$ with $k_i$ degrees of freedom. For each value of  $M \in \{1, 2, \ldots, 10\}$, we generate 256 independent RFI values such that $P_{RFI}^i \sim \chi^{2} (k_i)$ for $i=1,2,\ldots,256$ and $k_i \in\{1, 2, \ldots, M\}$. We then use these values to estimate $p_{soil}$ using SMAP's approach and our proposed weighted sum of samples approach.
\vspace{3pt}

Our primary requirement on the SMAP side is that we know the sample mean vector $\boldsymbol \mu$ and the covariance matrix $\boldsymbol \Sigma$, but we do not know the exact RFI realizations $P_{RFI}^{i} = p_{RFI}^{i}$.  
With the assumption that $P_{RFI}^i \sim \chi^{2} (k_i)$, this requirement is equivalent to knowing the number of active RFI sources $k_i$. Specifically, $\mu_{i} = \mathbb{E}(P_{RFI}^{i}|k_i)=k_i$ and $\sigma_{i,i}^{2}=2k_i$.
Since each RFI source impacts $p_i$ samples independently (equivalently, each $k_i$ is independent), the covariance terms $\sigma_{i,j}^{2}=0$ for all $i \neq j$. The mean vector $\boldsymbol \mu$ and covariance matrix $\boldsymbol \Sigma$ can be determined accordingly. 
\vspace{3pt}

As a simplified RFI detection technique similar to SMAP's, we use the general test of equation \eqref{eqq9} for all sample $p_i$ for $i=1,2,\ldots,256$.  Similar to SMAP, we average all the samples passing the test. Consequently, the estimator for SMAP is $\bar{p}=\frac{1}{g}\sum_{i=1}^{g} p_{i}$, where $g$ is the set of samples declared RF-free by the test in equation \eqref{eqq9}. In \eqref{eqq9}, we take $\beta=1$ and $m$ and $\sigma$ are respectively the arithmetic mean and standard deviation of all 256 $p_i$ samples.    

\vspace{3pt}

\begin{figure}
  \includegraphics[width=0.9\linewidth]{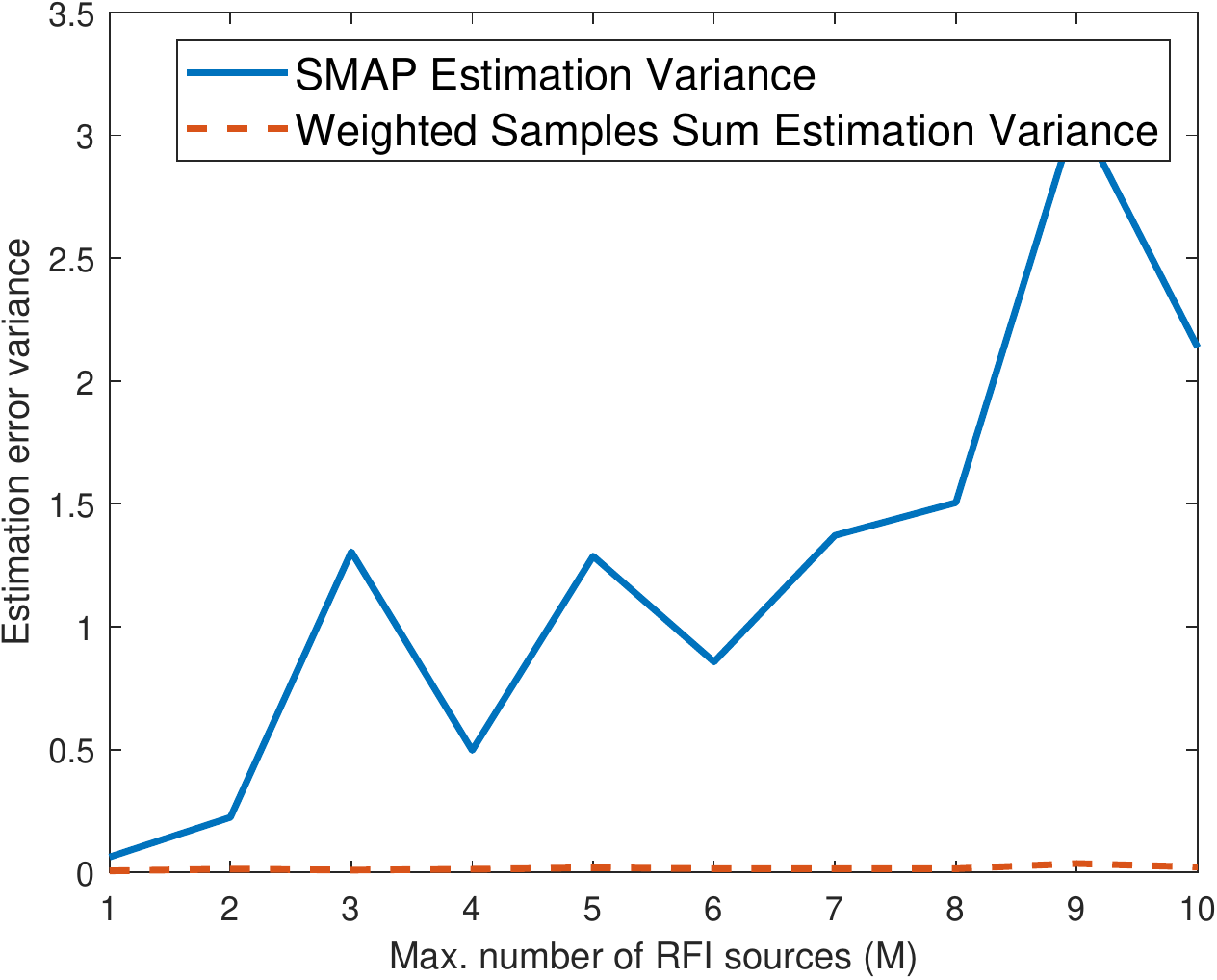}
  \caption{Estimation error variance.}
  \label{Var}
\end{figure}

Fig. \ref{Err} shows estimation error values for our proposed weighted sum of samples and SMAP's estimator. The horizontal axis represents $M$, i.e., the maximum number of RFI sources present in the environment. As observable in Fig. \ref{Err}, our weighted sum of samples method has a much lower estimation error. Moreover, it can preserve low and nearly constant estimation error in harsher RFI environments, while SMAP's estimation error increases approximately linearly in the number of interference sources. Fig. \ref{Var} shows estimation error variances for our proposed weighted sum of samples method and SMAP's estimator. Our weighted sum of samples method has a much lower estimation error variance, especially in harsher RFI environments.
\vspace{4pt}

\section{Conclusion \& Future Work}
\label{sec:con}
This paper proposes an optimized soil moisture estimator based on SMAP's digital backend design. Instead of  excluding RFI-contaminated samples and averaging the remaining ones, we investigate a probabilistic approach, where each sample is assigned a weight value in a weighted sample sum. The weights are determined through quadratic programming, using the mean values and covariance matrix of RFI. Simulations demonstrate this method's ability to make more accurate estimates. Assumptions are made regarding the knowledge of statistical properties such as the mean values and covariance matrix of the RFI distributions contaminating samples. Future work will investigate techniques for determining RFI probability distribution functions of samples, means, and covariance matrices. Additionally, real-world applications will be examined through a testbed incorporating a passive radiometer that is currently being calibrated and tested.
\bibliographystyle{IEEEtran}
\bibliography{refs.bib}

\end{document}